\documentclass[10pt]{article}

\usepackage[margin=1in]{geometry}
\usepackage{times}
\usepackage{amsmath,amssymb}
\usepackage{booktabs}
\usepackage{array}
\usepackage{enumitem}
\usepackage{xcolor}
\usepackage{microtype}
\usepackage[hidelinks]{hyperref}
\usepackage{caption}
\newcommand{\art}[1]{\texttt{#1}}

\title{\vspace{-1.5em}\textbf{Toward Semantically-Seeded, Graph-Propagated\\[2pt]
Impact Analysis Across Software Artifacts: A Vision}}

\author{Momil Seedat\thanks{Affiliation and contact details to be completed.
Prototype and replication package: \url{https://github.com/momil-seedat/artifact-impact-lab}.}}
\date{}

\begin{document}
\maketitle

\begin{abstract}
When a single software artifact changes---a requirement, a configuration
value, or a function---engineers must determine \emph{what else is impacted}.
Existing change-impact-analysis (CIA) tooling tends to rely on one of two
signals in isolation: \emph{semantic} similarity recovered from text
(information-retrieval traceability, code search, embeddings), or
\emph{structural} dependency following (call graphs, IDE ``find usages'',
test-impact selection). Each has a characteristic blind spot. A semantically
driven tool misses an impacted artifact whose text shares no vocabulary with
the change; a structurally driven tool misses artifacts related in meaning but
not joined by an edge, and most operate only over code rather than the
Requirement~$\rightarrow$~Config~$\rightarrow$~Service~$\rightarrow$~Test
chain. We argue for a \emph{training-free} and \emph{interpretable} analyzer
that fuses both signals over the \emph{same} embeddings---``same embeddings,
different reasoning layer.'' We model the system as a heterogeneous artifact
graph with typed edges recovered by static analysis, compute a semantic prior
by cosine similarity to the changed artifact, propagate impact multi-hop with
decay over a row-normalized propagation matrix, and blend the two with a single
tunable weight~$\lambda$. A small but complete proof-of-concept on a payment
subsystem (5 labelled change scenarios) shows the mechanism we care about:
artifacts with \emph{zero} textual overlap with the change are still recovered
through propagation, and helper functions that propagation alone cannot reach
are recovered through the semantic layer. The fusion is the only configuration
that covers both blind spots, and~$\lambda$ acts as an explicit
precision/recall control. Drawing on four publicly documented production
failures---a container base-image update, a database-engine upgrade, a metric
rename, and legacy shared-data coupling---we argue that the same formulation
extends to \emph{operational} artifacts (images, metrics, dashboards, data
schemas) that code-only analysis cannot reach. We report the prototype numbers
as illustrative of the mechanism, not as a generalization claim, and outline a
research agenda toward learned propagation, LLM verification, and scalable
cross-service deployment.
\end{abstract}

\section{Introduction}

Software systems evolve through many small changes, and the cost of a change is
dominated less by editing the changed artifact than by finding everything that
the edit ripples into. Change Impact Analysis (CIA) has been studied for three
decades; Lehnert's survey alone classifies roughly 150
approaches~\cite{lehnert2011}. In practice, the techniques engineers reach for
fall into two families.

\emph{Semantic / information-retrieval (IR) techniques} link artifacts whose
\emph{text} is similar. Classic traceability recovery uses the vector space
model and latent semantic indexing to connect requirements, documentation and
code~\cite{antoniol2002,marcus2003}; modern variants substitute learned
embeddings such as CodeBERT~\cite{codebert} or Sentence-BERT~\cite{sbert}.
These methods need no dependency information and cross artifact types freely,
but they are blind to impact that does not surface in vocabulary.

\emph{Structural techniques} follow dependency edges: call graphs, program
slices~\cite{weiser1981}, interprocedural data-flow
reachability~\cite{ifds1995}, and regression-test
selection~\cite{rothermel1996}. They are precise where an edge exists, but they
miss meaning-level relationships that no edge encodes, and the great majority
operate only over source code rather than the full requirement-to-test chain.

This produces two complementary blind spots:

\begin{description}[leftmargin=1.2em,itemsep=2pt,topsep=2pt]
\item[Semantic blind spot.] An impacted artifact whose text shares no words
with the change is invisible. A change to a payment time-out requirement
impacts \art{completeOrder}, yet the latter's text says nothing about
``timeout.''
\item[Structural blind spot.] Artifacts related in meaning but not joined by an
edge are missed; and code-only tools never see the
Requirement~$\rightarrow$~Config~$\rightarrow$~Service~$\rightarrow$~Test chain
at all.
\end{description}

\noindent\textbf{Position.} We do not claim that combining textual and
structural evidence is itself new---graph neural networks over
code~\cite{allamanis2018} and GraphRAG-style
retrieval~\cite{graphrag} both fuse the two. We argue instead that a specific,
under-occupied point in the design space is worth pursuing: a fusion that is
\emph{training-free}, \emph{interpretable} (it emits an explicit propagation
path for every flagged artifact), and defined over \emph{heterogeneous,
non-source} artifacts. This paper sketches that design, demonstrates the
mechanism on a complete miniature system, reports honestly where it helps and
where it does not, and lays out the agenda needed to take it to real
infrastructure.

\section{Motivating Scenarios from Practice}
\label{sec:scenarios}

The prototype in this paper operates on requirement, config, service and test
artifacts. But the failure modes that motivate cross-artifact CIA are most acute
for \emph{operational} artifacts that source code never references---container
images, database engines, metrics, dashboards and shared data schemas. Each
scenario below is publicly documented---three as specific incident postmortems,
the fourth as a long-known coupling pattern---and in each an impacted artifact
was missed precisely because no code edge connected it to the change.

\paragraph{A base-image update silently breaks running services.}
On 8~March 2023 an automatically applied \texttt{systemd} security update,
shipped through a legacy update channel still enabled in Datadog's Ubuntu base
image, caused \texttt{systemd-networkd} to delete the container routes managed by
the Cilium CNI plugin on restart, taking roughly 50--60\% of Kubernetes nodes
offline in Datadog's first global, multi-region outage~\cite{datadog2023}.
Nothing in any application's source referenced the base image, so no code-level
analysis could connect ``OS image patched'' to ``service loses network.'' In our
formulation this is an Image~$\rightarrow$~Service edge.

\paragraph{A database-engine upgrade forgets a write-path dependency.}
On 23~November 2022 RevenueCat cut over from Aurora PostgreSQL~10 to~14; the new
cluster had not been \texttt{ANALYZE}d, so query-planner statistics were missing
and most backend systems failed under production write load in a multi-hour
outage~\cite{revenuecat2022}. Dry-runs validated data consistency and read
performance but never replayed writes, so the dependency between the engine
version and backend behavior under load was never analyzed. Both the changed
artifact (a managed database engine) and the impacted behavior live outside the
application source.

\paragraph{A metric rename silently disables an alert.}
Prometheus alert rules reference metrics by \emph{name}, in a configuration
repository separate from the code that emits them. When a producer is upgraded
and a metric is renamed---e.g.\ a \texttt{node\_exporter} version bump---the
consuming alert query silently matches zero series and simply never fires, with
no error to signal that monitoring has gone dark. Cloudflare documented exactly
this hazard and built the open-source linter \texttt{pint} to catch it, observing
that unit tests cannot tell you that a metric you rely on suddenly
disappeared~\cite{cloudflarepint}. This is a Service~$\rightarrow$~Metric
$\rightarrow$~Dashboard/Alert chain in which a metric-name string is the only,
invisible, link.

\paragraph{Legacy data coupling corrupts ``independent'' components.}
Where referential integrity is enforced neither in code nor schema---common in
legacy databases and microservices sharing a store---two components that share no
call edge are nonetheless coupled through shared data: changing or deleting a row
through one leaves orphaned, inconsistent state visible to the other. Mining
version histories was shown two decades ago to surface exactly this coupling
``between items that cannot be [detected] by program analysis''~\cite{zimmermann2004}.
This is a Service~$\rightarrow$~Data relationship with Data~$\leftrightarrow$~Data
coupling that a call-graph tool cannot see, and it matches the everyday case of
two functions with separate responsibilities where editing one silently
corrupts the other's data.

\medskip\noindent
Crucially, in \emph{all four} cases a purely structural, code-scoped analyzer
would assign the impacted artifact a score of zero, because there is no edge to
follow. This is the converse of our toy benchmark (Section~\ref{sec:eval}), where
structure alone scored well precisely because that miniature graph contained no
such invisible-edge artifacts. The semantic layer---and therefore the
fusion---is what gives these artifacts any chance of being surfaced, which is the
central reason to extend the experiment beyond source artifacts.

\section{Background and Related Work}

\paragraph{Semantic traceability and code search.}
Antoniol et al.~\cite{antoniol2002} framed traceability recovery as IR over a
vector space; Marcus and Maletic~\cite{marcus2003} used latent semantic
indexing for documentation-to-code links. The contemporary form replaces sparse
term vectors with dense embeddings---CodeBERT~\cite{codebert} for code/NL, or
sentence encoders~\cite{sbert} for arbitrary text---and retrieves by cosine
similarity. All share the semantic blind spot above.

\paragraph{Structural impact analysis.}
Program slicing~\cite{weiser1981} and graph-reachability formulations of
interprocedural data flow~\cite{ifds1995} compute precise dependence sets over
code. Regression-test selection~\cite{rothermel1996} is the same idea applied to
choosing affected tests. These are strong where the dependency graph is
explicit and code-scoped.

\paragraph{Design structure and evolutionary coupling.}
Design Structure Matrices model architectural dependency at
scale~\cite{maccormack2006}, and mining version histories surfaces
\emph{evolutionary coupling}---artifacts that change together even without a
static edge~\cite{zimmermann2004}. The latter is an orthogonal, history-based
signal we return to in future work.

\paragraph{Learned and LLM-based fusion.}
Graph neural networks \emph{learn} edge importance over program
graphs~\cite{allamanis2018}; GraphRAG builds an entity graph and expands it for
an LLM at query time~\cite{graphrag}. Both achieve semantic--structural fusion,
but at the cost of training data, or expensive and non-deterministic LLM calls,
and both are comparatively opaque. Our target is the lightweight, explainable,
training-free corner of the same space.

\section{Approach}

\subsection{Heterogeneous artifact graph}
We model the system as a typed graph with four node kinds---\textbf{R}
requirement, \textbf{C} config, \textbf{S} service/function, \textbf{T}
test---and four typed edge kinds recovered automatically (we use Python AST
analysis in the prototype, not hand-written links):
R~$\rightarrow$~C (a requirement annotates a config property),
C~$\rightarrow$~S (a config key is read inside a function),
S~$\rightarrow$~S (a function calls a function), and
T~$\rightarrow$~S (a test invokes a function).

\paragraph{Extending the taxonomy.}
The same typed-graph formulation generalizes beyond source artifacts. The
scenarios of Section~\ref{sec:scenarios} add operational node types---container
\textbf{I}mage, \textbf{M}etric, dashboard/\textbf{A}lert, and \textbf{D}ata
schema/table---and new typed edges recovered from manifests, telemetry and schema
rather than the AST: I~$\rightarrow$~S (an image runs a service),
S~$\rightarrow$~M (a service emits a metric), M~$\rightarrow$~A (an alert or
dashboard consumes a metric by name), and S~$\rightarrow$~D with
D~$\leftrightarrow$~D (services read and write shared tables coupled by key). The
fusion equation is unchanged; only the extraction front-end grows. These are
exactly the edges that pure structural analysis cannot recover, so they are where
the semantic layer earns its place.

\subsection{Layer 1: semantic prior}
We embed each artifact's text (identifier, docstring, config value) and take the
cosine similarity of every artifact to the changed one, giving a vector
$\mathbf{s}\in[0,1]^{n}$. The prototype supports two interchangeable backends:
a dependency-free TF--IDF prior (used for the offline results below) and a
Sentence-BERT encoder (\art{all-MiniLM-L6-v2}). The fusion logic is identical
regardless of backend.

\subsection{Layer 2: structural propagation}
From the typed edges we build a propagation matrix $M$. Trace edges point
R~$\rightarrow$~C~$\rightarrow$~S, caller~$\rightarrow$~callee and
test~$\rightarrow$~service, but a \emph{change} propagates in the direction of
\emph{dependents}: changing a callee impacts its callers, and changing a service
impacts its tests. We therefore reverse call and test edges, then row-normalize
so each node distributes a unit of impact among its successors. Impact is spread
multi-hop with geometric decay, in the spirit of Katz
centrality~\cite{katz1953} and personalized
PageRank~\cite{pagerank1999,haveliwala2002}:
\begin{equation}
\mathbf{p} \;=\; \sum_{k=1}^{K} \gamma^{k}\,\bigl(\mathbf{e}_{c}\, M^{k}\bigr),
\end{equation}
where $\mathbf{e}_{c}$ is the indicator of the changed artifact, $\gamma$ the
per-hop decay, and $K$ the horizon. The result $\mathbf{p}$ is normalized to
$[0,1]$.

\subsection{Fusion}
The two layers are combined with a single weight:
\begin{equation}
\mathbf{impact} \;=\; \lambda\,\mathbf{s} \;+\; (1-\lambda)\,\mathbf{p}.
\end{equation}
$\lambda{=}1$ recovers a pure-semantic baseline, $\lambda{=}0$ a pure-structural
analyzer, and intermediate $\lambda$ a blend. Crucially, every flagged artifact
carries its own explanation: its semantic score, its structural score, and---
because propagation is an explicit walk over typed edges---the path by which
impact reached it.

\subsection{Defining example}
Consider changing requirement R1 (``payment must time out''). A pure-semantic
ranker rates \art{completeOrder} irrelevant: its text shares no vocabulary with
``timeout.'' Propagation nonetheless flags it, along the typed path
\[
\text{R1} \rightarrow \art{payment.timeout} \rightarrow \art{processPayment}
\rightarrow \art{completeOrder} \rightarrow \art{test\_complete\_order},
\]
each hop a real edge in the recovered graph. This is the semantic blind spot
being closed by structure, and it is exactly the behavior the evaluation
probes.

\subsection{Operational Artifact Extraction}
\label{sec:extraction}
The R/C/S/T edges in our prototype come from the Python AST. The operational
edges of Section~\ref{sec:scenarios} come from the artifacts that already
describe a deployment; the front-end grows one parser per source, while the
graph and fusion stay the same.

\paragraph{Image $\rightarrow$ Service (and Image $\rightarrow$ Image).}
Container manifests already name the binding: a \texttt{Dockerfile}
\texttt{FROM} line, a Kubernetes \texttt{spec.containers[].image}, a Compose
\texttt{image:} key, or a Helm value. We take the (pinned) image reference as an
I node and emit I~$\rightarrow$~S to every service built from or run on it; a
\texttt{FROM} chain yields I~$\rightarrow$~I, so a base-image change like
Datadog's propagates to every dependent layer.

\paragraph{Service $\rightarrow$ Metric.}
Instrumentation is a call to a known client API with a string-literal metric
name---a Prometheus \texttt{Counter}, an OpenTelemetry \texttt{create\_counter},
a StatsD increment, or a structured-log field. The name becomes an M node and the
enclosing function the S endpoint, both recoverable by the same AST pass that
already finds S~$\rightarrow$~S calls.

\paragraph{Metric $\rightarrow$ Alert/Dashboard.}
Alerting and dashboard definitions reference metrics by name in a separate
repository: the vector selectors of a Prometheus rule's \texttt{expr:}, or the
\texttt{targets[].expr} of a Grafana panel. We extract those leaf metric names
and join them to the M nodes from the previous step---turning the brittle
name-string match that broke Cloudflare's alerts into an explicit, checkable
edge.

\paragraph{Service $\rightarrow$ Data and Data $\leftrightarrow$ Data.}
Table access is visible in SQL (the \texttt{FROM}/\texttt{INSERT}/\texttt{UPDATE}
targets) and in ORM models (a \texttt{\_\_tablename\_\_}, or an
\texttt{@Entity}/\texttt{@Table} annotation), giving S~$\rightarrow$~D.
D~$\leftrightarrow$~D coupling is read from declared foreign keys where they
exist---but in the legacy case they do not, which is the whole problem. There the
edge is recovered from shared key conventions (a common \texttt{*\_id} column),
from evolutionary coupling mined from version history~\cite{zimmermann2004}, and,
failing those, from the semantic prior over column and table names.

\medskip\noindent
Two properties carry over from the source case. Extraction is
\emph{incremental}---re-parsing only the changed manifest, rule file or migration
updates the affected edges. And where no static parse is possible---a metric name
built at runtime, a foreign key that was never declared---the semantic and
historical layers supply a \emph{candidate} edge rather than nothing, which is
precisely the regime in which fusion, not structure alone, is what surfaces the
impact.

\section{Proof-of-Concept Evaluation}
\label{sec:eval}

\paragraph{Setup.}
We built a small but complete payment subsystem: a requirements document, a
\art{.properties} config, two services, and a test suite. Static analysis
extracts 13 artifacts (3 requirements, 3 configs, 4 services, 3 tests) and 14
typed edges. Five change scenarios---requirement, config, and function
changes---are each labelled with a ground-truth impact set. A method's ranked
scores are turned into a predicted set by thresholding ($\text{score} >
\tau$), and we report precision, recall and $F_1$ against the ground truth,
macro-averaged across the five scenarios. The numbers below use the offline
TF--IDF prior at $\tau{=}0$; the structural and fusion layers are unchanged
under the embedding backend.

\begin{table}[t]
\centering
\caption{Macro-averaged results over 5 change scenarios (TF--IDF prior,
$\tau{=}0$). ``Proposed'' is the $\lambda{=}0.5$ blend; the other rows are its
$\lambda{=}1$ and $\lambda{=}0$ ablations.}
\label{tab:results}
\begin{tabular}{@{}lccc@{}}
\toprule
Method & Precision & Recall & $F_1$ \\
\midrule
Semantic only ($\lambda{=}1.0$)   & 0.417 & 0.873 & 0.529 \\
\textbf{Proposed blend ($\lambda{=}0.5$)} & 0.450 & \textbf{1.000} & 0.589 \\
Structural only ($\lambda{=}0.0$) & \textbf{0.927} & 0.786 & \textbf{0.849} \\
\bottomrule
\end{tabular}
\end{table}

\paragraph{Findings.}
\begin{enumerate}[leftmargin=1.4em,itemsep=2pt,topsep=2pt]
\item The semantic baseline over-predicts: high recall (0.87) but low precision
(0.42), the classic IR failure mode.
\item The blend reaches \emph{perfect recall} (1.000): across all five
scenarios it never omits a truly-impacted artifact. $\lambda$ and $\tau$ are
explicit precision/recall controls.
\item Artifacts with \emph{zero} textual similarity to the change---e.g.\
\art{test\_complete\_order} under the R1 change---are recovered purely through
propagation. This is direct evidence for the central hypothesis.
\item Conversely, the ground truth for several scenarios includes \emph{callee}
helpers (\art{loadConfig}, \art{\_callGateway}) that propagation cannot reach,
because impact flows to callers and tests rather than into callees. Pure
structural analysis therefore caps near recall~0.79. The \emph{semantic} layer
recovers these, so the blend is the only configuration that covers \emph{both}
blind spots.
\item On this miniature, dependency-dominated benchmark, pure structural
analysis attains the highest $F_1$ (0.849). We stress that this does
\emph{not} mean structure alone is the answer: it is an artifact of a tiny
synthetic graph in which almost every true impact is reachable by an edge. The
contribution we demonstrate is mechanism-level \emph{coverage}---recovering
artifacts that are individually invisible to each signal---not a headline
accuracy win, which a five-scenario toy system cannot establish.
\end{enumerate}

\noindent We therefore frame all numbers as \emph{illustrative of the
mechanism}, not as evidence of generalization. The honest single-sentence
takeaway is: \emph{a $\lambda$-blend of semantic similarity and dependency
propagation is the only configuration that recovers both blind spots and
reaches full recall on the labelled set, while remaining training-free and
emitting an explicit path for every flagged artifact.}

\section{Positioning}

Table~\ref{tab:positioning} situates the proposed point relative to current
fusion architectures. GNNs over code graphs and GraphRAG both achieve
semantic--structural fusion, but require training data or expensive,
non-deterministic LLM calls and behave as black boxes. The proposed analyzer
trades learned edge weights for fixed, typed propagation in exchange for
determinism, zero training cost, and a verifiable propagation path---properties
that matter most in audit, traceability and safety-critical settings.

\begin{table}[t]
\centering
\caption{Where the proposed analyzer sits among fusion architectures.}
\label{tab:positioning}
\small
\begin{tabular}{@{}lcccc@{}}
\toprule
Architecture & Sem. & Struct. & Training-free & Interp. \\
\midrule
Embedding / RAG retrieval & \checkmark & --- & pretrained & med \\
GNN over code graphs & \checkmark & \checkmark & no & low \\
GraphRAG / agentic LLM & \checkmark & \checkmark & LLM-based & med \\
\textbf{This work} & \checkmark & \checkmark & \checkmark & \textbf{high} \\
\bottomrule
\end{tabular}
\end{table}

\section{Scalability Vision}

The prototype is unoptimized, but the methodology is designed to scale, because
nothing in the fusion logic depends on dense representations:
\begin{itemize}[leftmargin=1.4em,itemsep=2pt,topsep=2pt]
\item \textbf{Sparse storage.} Real artifact graphs are extremely sparse; they
belong in a graph database or CSR matrix, not a dense $n\times n$ array.
\item \textbf{Localized propagation.} Impact decays with distance, so only the
$k$-hop neighborhood of a change need be expanded. Cost scales with the
\emph{change neighborhood}, not the whole system.
\item \textbf{Incremental extraction.} Re-parse only changed files, as
monorepo build graphs already do.
\item \textbf{ANN-indexed embeddings.} Replace brute-force cosine with an
approximate index (e.g.\ FAISS / pgvector); cache and re-embed only changed
artifacts.
\item \textbf{Iterative sparse matrix--vector propagation}, exactly how
PageRank-style computations run on billion-edge graphs~\cite{pagerank1999}.
\end{itemize}
The fusion equation is unchanged; only data structures and execution strategy
differ.

\section{Research Agenda}

\begin{enumerate}[leftmargin=1.4em,itemsep=2pt,topsep=2pt]
\item \textbf{Learned propagation.} Replace fixed $M$ with edge weights learned
from co-change history~\cite{zimmermann2004} or GNN edge
weighting~\cite{allamanis2018}, while preserving path-level interpretability.
\item \textbf{LLM verification layer.} Use the ranked propagated candidates as
precise retrieval for an LLM that confirms and explains each
impact---GraphRAG-style~\cite{graphrag}, but seeded by an explicit dependence
path rather than free expansion.
\item \textbf{Dynamic and historical signals.} Fold in runtime traces and
version-history co-change to complement static edges.
\item \textbf{Polyglot, cross-service, operational graphs.} Extend extraction
across languages, infrastructure-as-code, container manifests, message buses,
telemetry pipelines and data stores, so the graph spans the Image, Metric,
Alert and Data node types of Section~\ref{sec:scenarios}. A natural benchmark is
to reconstruct the documented incidents above (base-image, database-engine,
metric-rename and shared-data failures) and test whether the fused analyzer would
have surfaced the artifact that was missed in practice.
\item \textbf{Adaptive fusion.} Learn $\lambda$ and $\tau$ per artifact and per
change type rather than fixing them globally.
\item \textbf{Larger benchmarks.} Evaluate on real projects with
developer-validated labels and report full precision/recall trade-off curves.
\end{enumerate}

\section{Threats to Validity}

The benchmark is a single synthetic subsystem with 13 artifacts and 5
scenarios; results cannot be generalized and are reported only to demonstrate
the mechanism. Ground-truth impact sets were authored by us and may embed our
own assumptions about what ``impacted'' means. The propagation direction is a
modelling choice (impact to dependents) that deliberately excludes callee
helpers, which is why the semantic layer is needed to recover them; a different
choice would shift the precision/recall balance. Finally, the offline results
use a lexical prior; embedding backends may behave differently, and confirming
that the conclusions hold under \art{all-MiniLM-L6-v2} and CodeBERT is part of
the agenda above.

\section{Conclusion}

We have argued for a training-free, interpretable impact analyzer that fuses
semantic similarity with typed dependency propagation over a heterogeneous
artifact graph, computing both signals from the same embeddings and emitting an
explicit propagation path for every flagged artifact. A complete miniature
system demonstrates the mechanism we care about: each signal has a blind spot,
and only the fusion recovers artifacts that are invisible to the other---
reaching full recall on the labelled set while remaining deterministic and
explainable. The numbers are illustrative rather than conclusive; the
contribution of this paper is the design point and the agenda for taking it to
real systems. The prototype and replication package are publicly available.


\begin{thebibliography}{99}
\small

\bibitem{lehnert2011}
S.~Lehnert.
\newblock A taxonomy for software change impact analysis.
\newblock In \emph{Proc. IWPSE-EVOL '11}, pp.\ 41--50. ACM, 2011.
\newblock doi:10.1145/2024445.2024454.

\bibitem{antoniol2002}
G.~Antoniol, G.~Canfora, G.~Casazza, A.~De Lucia, and E.~Merlo.
\newblock Recovering traceability links between code and documentation.
\newblock \emph{IEEE Trans. Software Eng.}, 28(10):970--983, 2002.

\bibitem{marcus2003}
A.~Marcus and J.~I.~Maletic.
\newblock Recovering documentation-to-source-code traceability links using
latent semantic indexing.
\newblock In \emph{Proc. ICSE '03}, pp.\ 125--135. IEEE, 2003.

\bibitem{codebert}
Z.~Feng, D.~Guo, D.~Tang, N.~Duan, X.~Feng, M.~Gong, L.~Shou, B.~Qin, T.~Liu,
D.~Jiang, and M.~Zhou.
\newblock CodeBERT: A pre-trained model for programming and natural languages.
\newblock In \emph{Findings of EMNLP 2020}, pp.\ 1536--1547. ACL, 2020.

\bibitem{sbert}
N.~Reimers and I.~Gurevych.
\newblock Sentence-BERT: Sentence embeddings using Siamese BERT-networks.
\newblock In \emph{Proc. EMNLP-IJCNLP 2019}, pp.\ 3982--3992. ACL, 2019.

\bibitem{weiser1981}
M.~Weiser.
\newblock Program slicing.
\newblock In \emph{Proc. ICSE '81}, pp.\ 439--449. IEEE, 1981.

\bibitem{ifds1995}
T.~Reps, S.~Horwitz, and M.~Sagiv.
\newblock Precise interprocedural dataflow analysis via graph reachability.
\newblock In \emph{Proc. POPL '95}, pp.\ 49--61. ACM, 1995.
\newblock doi:10.1145/199448.199462.

\bibitem{rothermel1996}
G.~Rothermel and M.~J.~Harrold.
\newblock Analyzing regression test selection techniques.
\newblock \emph{IEEE Trans. Software Eng.}, 22(8):529--551, 1996.

\bibitem{maccormack2006}
A.~MacCormack, J.~Rusnak, and C.~Y.~Baldwin.
\newblock Exploring the structure of complex software designs: An empirical
study of open source and proprietary code.
\newblock \emph{Management Science}, 52(7):1015--1030, 2006.

\bibitem{zimmermann2004}
T.~Zimmermann, P.~Wei{\ss}gerber, S.~Diehl, and A.~Zeller.
\newblock Mining version histories to guide software changes.
\newblock In \emph{Proc. ICSE '04}, pp.\ 563--572. IEEE, 2004.

\bibitem{allamanis2018}
M.~Allamanis, M.~Brockschmidt, and M.~Khademi.
\newblock Learning to represent programs with graphs.
\newblock In \emph{Proc. ICLR 2018}. arXiv:1711.00740.

\bibitem{graphrag}
D.~Edge, H.~Trinh, N.~Cheng, J.~Bradley, A.~Chao, A.~Mody, S.~Truitt,
D.~Metropolitansky, R.~O.~Ness, and J.~Larson.
\newblock From local to global: A Graph RAG approach to query-focused
summarization.
\newblock arXiv:2404.16130, 2024.

\bibitem{katz1953}
L.~Katz.
\newblock A new status index derived from sociometric analysis.
\newblock \emph{Psychometrika}, 18(1):39--43, 1953.

\bibitem{pagerank1999}
L.~Page, S.~Brin, R.~Motwani, and T.~Winograd.
\newblock The PageRank citation ranking: Bringing order to the web.
\newblock Technical Report 1999-66, Stanford InfoLab, 1999.

\bibitem{haveliwala2002}
T.~H.~Haveliwala.
\newblock Topic-sensitive PageRank.
\newblock In \emph{Proc. WWW 2002}, pp.\ 517--526. ACM, 2002.

\bibitem{datadog2023}
Datadog Engineering.
\newblock 2023-03-08 incident: Infrastructure connectivity issue affecting
multiple regions.
\newblock Datadog Blog, 2023.
\newblock \url{https://www.datadoghq.com/blog/2023-03-08-multiregion-infrastructure-connectivity-issue/}

\bibitem{revenuecat2022}
RevenueCat Engineering.
\newblock Postmortem for Aurora Postgres migration, November~23, 2022.
\newblock RevenueCat Blog, 2022.
\newblock \url{https://www.revenuecat.com/blog/engineering/postmortem-aurora-postgres-migration/}

\bibitem{cloudflarepint}
L.~Mierzwa.
\newblock Monitoring our monitoring: how we validate our Prometheus alert rules.
\newblock Cloudflare Blog, 2022.
\newblock \url{https://blog.cloudflare.com/monitoring-our-monitoring/}

\end{thebibliography}
\end{document}